\documentclass[preprint,12pt]{elsarticle}  
\usepackage{amsmath}
\usepackage{amssymb}                                          
\usepackage{graphicx}

\usepackage{natbib}
\usepackage{mathtext}
\usepackage{epsf}
\usepackage{amsfonts}
\usepackage{gensymb}
\usepackage{float}

\usepackage[T2A]{fontenc}    
\usepackage[utf8]{inputenc}  

\title{FRBs and magnetar activity statistics}
\author[1]{Sergei B. Popov}
\affiliation[1]{ICTP -- Abdus Salam International Center for Theoretical Physics }

\begin{document}
\maketitle

\section*{Abstract}
With simple estimates based on recent observational data and the assumption that among known FRB sources most repeaters with extremely high rates of repetition (super-repeaters) are already identified, we demonstrate that the hypothesis that super-repeaters and one-off events come from the same population of magnetars is not self-contradictory. 
In this toy model, the super-repeater stage has a duration of about a few years and the period when one-off events are mostly emitted corresponds to the active life of a magnetar $\sim $few$\times 10^3$~years. Intervals between strong events (observed as one-off FRBs) from the same source are $\sim10$~years, corresponding to the expected time interval between giant flares.

\pagebreak 

\section{Introduction}

Fast radio bursts (FRBs) -- millisecond-scale extragalactic radio bursts, -- are still not a well-understood phenomenon, see a detailed recent review in \cite{2022arXiv221203972Z}.
 At the moment, after $\Delta t \sim$ few yrs of observations, \footnote{Here the CHIME monitoring is mostly important, see e.g. \cite{2023arXiv230705261L} where the authors based on observations in far side-lobes put a lower limit $\gtrsim 10^4$~hours for repetitions of one-off events. Still, $\Delta t$ can be also taken $\sim 10$~yrs, but in this short research note, I provide estimates by an order of magnitude. }  we have about a thousand one-off FRBs ($N_{one}^{obs} \sim  10^3$) \cite{2021ApJS..257...59C}  and $> 50$ repeaters \cite{2023ApJ...947...83C, 2023arXiv230800336X}. Among repeaters, there are 5-6 sources with enormous rates of bursts ($N_{rep}^{obs}\sim$ few):
 FRBs 20121102A,
20180916B, 20190520B, 20201124, see e.g. \cite{2022arXiv221205242H}; FRB 20220912A \cite{2023arXiv230414665Z}; and FRB 20200120 \cite{2023MNRAS.520.2281N}. Let us call them ``super-repeaters''.
Recently it was shown that due to high energy release in super-repeaters their life cycle, $T_{rep}$, cannot be much longer than a few years \cite{2023arXiv230414665Z}. To make the estimate these authors assume that the total energy of a burst can be calculated from the isotropic radio energy as $E_{tot}=10^3 \times E_{radio}$. Note, that the value of $E_{tot}$ can be two orders of magnitude larger if one uses the data on the Galactic FRB from SGR 1935+2154 \cite{2021NatAs...5..372R} and neglect the beaming of the radio signal. Such high energy budget can conflict with the total amount of magnetic energy $E_{mag}\sim 2 \times 10^{47} \, B_{15}^2 R_6^3$~erg, where the normalized field is $B_{15}=B/10^{15}$~G and the normalized NS radius is $R_6=R/10^6$~cm, accounting for the observed period of activity of the known super-repeaters.

Under the standard hypothesis that FRB sources are active magnetars, the life cycle of one-off sources is expected to be about: $T_{one}\sim $few$\times  10^3$~yrs  (see a review, e.g. in \cite{2015RPPh...78k6901T}). During this period they can produce $N_b$ bursts. 

In a simple model, we can estimate $N_b$ under the assumption that super-repeaters and one-off events come from the same sources, but super-repeaters correspond to a distinct stage of activity of each object.
 Also, we can explain why presently for FRBs we have:

\begin{equation}
\frac{N_{one}^{obs}}{T_{one}}\approx\frac{N_{rep}^{obs}}{T_{rep}}.
\end{equation}

Comparison with the expected (from the Galactic magnetar statistics) number $N_b$ and with observed ratios $\frac{N_{one}^{obs}}{T_{one}}$ and $ \frac{N_{rep}^{obs}}{T_{rep}}$ can probe basic assumptions of the toy model.


\pagebreak

\section{Number of bursts}

Let us define $N_{one}$ -- the number of NSs in the sampled volume (where we can detect FRBs) which were/are sources of one-off FRBs during the Hubble time $T_{Hubble}\sim 10^{10}$~yrs (we can choose instead any time interval $\gg T_{one}$). 
Then, the number of observed one-off FRBs during $\Delta t$ is:

\begin{equation}
N_{one}^{obs} = \left(N_{one} \times \frac{T_{one}}{T_{Hubble}}\right) \times \frac{\Delta t}{T_{one}}\times N_b. 
\end{equation}

Similarly,

\begin{equation}
N_{rep}^{obs} = \left(N_{rep} \times \frac{T_{rep}}{T_{Hubble}}\right), \, \mathrm{if \, T_{rep} > \Delta t},
\end{equation}

and

\begin{equation}
N_{rep}^{obs} = \left( N_{rep} \times \frac{\Delta t}{T_{Hubble}}\right), \, \mathrm{if \, T_{rep} < \Delta t}.
\end{equation}

Below we use eq.(4) for the estimates, but again we underline that $T_{rep}\approx\Delta t$.

Let us define the formation rate of sources as $\nu=N/T_{Hubble}$.
 If we assume that super-repeaters are found only among already detected one-off FRBs in the whole volume, i.e. that $N_{rep}=N_{one}$
then, Obviously:

\begin{equation}
        \frac{\nu_{one}}{\nu_{rep}}= 1.
\end{equation}

Substituting eqs. (1-2) into eq. (5) and assuming $N_{one}=N_{rep}$ we obtain:

\begin{equation}
    N_b\approx \frac{N_{one}^{obs}}{N_{rep}^{obs}}\approx \mathrm{few} \times 10^2.
\end{equation}

Powerful radio flares observed as one-off events might be related to giant flares of extragalactic magnetars \cite{2007arXiv07102006P}. 
The interval between subsequent flares is $T_{one}/N_b\sim 10$~yrs, roughly compatible with known properties of young active magnetars. The magnetic energy budget of a magnetar is in correspondence with this total number of flares.  
So, the hypothesis that one-off events and super-repeaters form a single population is not self-contradictory.

\pagebreak

\section{Number of sources vs. time scale  ratios}

Now lets us discuss the approximate equality 
$\frac{N_{one}^{obs}}{T_{one}}\approx\frac{N_{rep}^{obs}}{T_{rep}}$
which appears from the present observational data. 
It can be naturally obtained under the assumptions we made: the super-repeater activity is a short stage of nearly all one-off sources and most of the presently active super-repeaters among detected sources of one-off events are identified. Also, we assume that super-repeaters are identified only among already detected sources, i.e. it is necessary to register the first strong burst to identify later a super-repeater. 


We have $N_{one}^{obs}$ one-off events. Their sources have lifetime $T_{one}$. The probability to catch any of these sources at a given moment at the stage of super-repeater is just $T_{rep}/T_{one}$. So, we immediately come to:

\begin{equation}
\frac{N_{one}^{obs}}{T_{one}}\approx\frac{N_{rep}^{obs}}{T_{rep}}.
\label{eq}
\end{equation}
This expression is valid till $\Delta t$ is not significantly larger than $T_{rep}$. 
As $\Delta t$ is growing, we must substitute $T_{rep}$ by this quantity in eq.~\ref{eq}. 
Already now the ratios are close to each other if we take $\Delta t$ instead of $T_{rep}$.
 But note, that $N_{one}$ can be larger than $\sim10^3$ as not all detections are reported, yet, and the rate of FRB discovery at CHIME is about a few events per day.


\pagebreak

\section{Discussion}



In the framework of the toy model described above we can expect that on the time scale of a few years presently known super-repeaters might exit their active phase, but new sources will appear. Also, we can expect that the approximate equality $\frac{N_{one}^{obs}}{T_{one}}\approx\frac{N_{rep}^{obs}}{\Delta t}$ will be valid as soon as new super-repeaters are effectively identified.

 Under the assumptions we made, the energy released during the active phase (super-repeater) is comparable to the energy released during the whole lifetime of a magnetar, as they are roughly limited by the amount of the magnetic energy $\sim (10^{47}-10^{48})\,  B_{15}^2 $~erg. Thus, it is not very probable that there are many episodes of super-repeater activity. Otherwise, it is necessary to assume that $T_{rep}$ of each episode is smaller (and only altogether they correspond to the time interval equal to $T_{rep}$) in contradiction with observations. Episodes of super-repeater activity might be related to the violent reconfiguration of a magnetic field in the crust of a (probably young) magnetar.


In the case of super-repeater FRB 20201124A, it is shown that for the brightest bursts the energy distribution is rather flat \cite{2023arXiv230615505K}. Thus, if a source entered the super-repeater stage, it would be quite probable to detect more than one strong (one-off-like) bursts from it which might trigger additional observations and finally -- identification of a super-repeater. So, we accept that the hypothesis that mostly presently active super-repeaters are already identified might be correct.

Here we neglected many selection effects (and, no doubt, detailed population synthesis calculations are necessary), but we think that such a toy-model approach is valid as a zero approximation and is very illustrative. If the basic assumptions that super-repeaters can be identified mostly among sources already registered as one-off events and that the majority of super-repeaters are already identified among known one-off sources are shown to be wrong, then automatically, the main conclusion is falsified (i.e., super-repeaters are a separate population).

\pagebreak

\section{Conclusions}

We conclude that the model where one-off FRBs and super-repeaters come from the same sources, but super-repeaters represent just a short stage in the history of the magnetar activity, is self-consistent and compatible with general properties of the magnetar activity and present-day statistics. 

\section*{Acknowledgements}
I thank drs. Maxim Barkov and Maxim Pshirkov for discussions.

\pagebreak


\bibliography{sample}

\end{document}